# Modeling approach to regime shifts of primary production in shallow coastal ecosystems

J. M. Zaldívar[1], F.S. Bacelar[†], S. Dueri, D. Marinov, P. Viaroli[§] and E. Hernández-García[†]

European Commission, Joint Research Centre, Via E. Fermi 2749,
21020-Ispra (VA), Italy

[†]IFISC, Instituto de Física Interdisciplinar y Sistemas Complejos (CSIC-UIB)
Campus Universitat de les Illes Balears, E-07122 Palma de Mallorca, Spain

[§]Department of Environmental Sciences,
University of Parma, Viale G. P. Usberti 33/A, 43100-Parma, Italy

**Abstract.**

Pristine coastal shallow systems are usually dominated by extensive meadows of seagrass species, which are assumed to take advantage of nutrient supply from sediment. An increasing nutrient input is thought to favour phytoplankton, epiphytic microalgae, as well as opportunistic ephemeral macroalgae that coexist with seagrasses. The primary cause of shifts and succession in the macrophyte community is the increase of nutrient load to water; however temperature plays also an important role. A competition model between rooted seagrass (*Zostera marina*), macroalgae (*Ulva sp*), and phytoplankton has been developed to analyse the succession of primary producer communities in these systems. Successions of dominance states, with different resilience characteristics, are found when modifying the input of nutrients and the seasonal temperature and light intensity forcing.

---

[1] Corresponding author. E-mail: jose.zaldivar-comenges@jrc.it; Institute for Health and Consumer Protection, TP272

# 1. INTRODUCTION

Shallow transitional water systems (McLusky and Elliot. 2007) are intrinsically unstable and highly variable over wide temporal and spatial scales (Kjerfve, 1994; Zaldívar et al., 2008). These ecosystems, being interfaces between terrestrial and aquatic ecosystems, provide essential ecological functions influencing the transport of nutrients, material and energy from land to sea (Wall et al., 2001). Biodiversity can attain low values, but its functional significance remains high (Sacchi, 1995). Therefore, shifts in diversity are likely to have important and profound consequences for ecosystem structure and functioning (Levin et al., 2001). Invasions, competitive advantages, and nonlinear feedback interactions may lead to alternating states and regime shifts (Scheffer et al., 2001) which, once occurred, may pose limits for remediation strategies since it may be difficult, if not impossible, returning to the original state (Webster and Harris, 2004).

Contrary to open seas, where primary production is dominated by phytoplankton, in transitional waters a considerable portion of primary production is performed by angiosperms, epiphytic algae, macroalgae and epibenthic microalgae. In addition, shallow aquatic ecosystems do not show the typical correlation between nutrient inputs and chlorophyll-a in water (Nixon et al., 2001), as already demonstrated for deeper coastal waters and lakes (Vollenweider, 1976).

Regime shift phenomena occurring in shallow coastal areas (Sand-Jensen and Borum, 1991; Nienhuis, 1992; Viaroli et al., 1996; Flindt et al., 1999; Schramm, 1999) have been already documented as the result of the competition between free floating plants and submerged phanerogams.

Several authors have proposed a conceptual scheme that considers nutrient inputs as the main driver in the succession from benthic vegetation to phytoplankton or floating seaweeds in shallow transitional waters (Nienhuis, 1992; Borum, 1996; Valiela et al., 1997; Hemminga, 1998; Nixon et al., 2001). This conceptual scheme is based on the evolution of benthic communities through several phases as the level of nutrients increases. In the pristine stage, the community is dominated by phanerogams species from a relatively small number of genera, i.e. *Zostera, Thalassia, Halodule, Cymodocea,* and *Ruppia.* Nutrient enrichment leads to an increase in epiphytic microalgae, followed by the increase in floating ephemeral macroalgae -as *Ulva* and *Gracilaria*- which compete for light and nutrients thus producing the disappearance of perennial seagrass species. Finally, at high levels of nutrient input, phytoplankton growth increases water turbidity enough to depress macroalgal growth thus leading to a dominance of phytoplankton species.

Even though the decline of seagrasses due to anthropogenic eutrophication is a worldwide phenomenon (Orth et al., 2006; Short et al., 2006), there is no direct causality evidence from field data (Ralph et al., 2006). In addition, it is not evident from experimental studies where the limit lies for the dominance shift between these two competing type of organisms are (Schramm, 1999; Hauxwell and Valiela, 2004). Field studies demonstrate that the decrease of seagrass meadows is directly related to nitrogen loadings (Nelson et al., 2003; Hauxwell and Valiela, 2004) and the dominance of macroalgae, especially Ulvaes, becomes apparent in eutrophic environments (Borum, 1996). An overview of seagrass responses to nutrient enrichment and/or eutrophication events is presented in Burkholder et al. (2007), whereas the evolution of several Mediterranean coastal lagoons from pristine conditions to the present situation is summarised in Viaroli et al. (2008).

Although nutrient loading is one of the main drivers of regime shifts in transitional waters, light and temperature have been also recognised as key abiotic factors controlling algal growth (Schramm, 1999). Furthermore, in transitional water ecosystems hydrological and hydrodynamic conditions affect community persistence (Dahlgreen and Kautsky, 2004; Marinov et al., 2007).

Regime shifts occurring in shallow aquatic ecosystems were analysed by Scheffer et al. (2003), who developed a minimum model with two ordinary differential equations, one considering floating plants and the other for submerged aquatic vegetation (SAV). Competition between floating vegetation and SAV was found to cause alternate attractors, since floating plants outcompete SAV when light is the only limiting factor, whereas SAV species dominate at low nutrient concentrations since they are able to uptake nutrients from the sediments. The model, although not fully validated with experimental data, was the first to provide a comprehensive explanation of several observed phenomena.

In this work, we have studied the regime shifts from SAV to floating macroalgae in shallow brackish ecosystems. We have developed a basic model that accounts for the competition between *Zostera* *marina* and *Ulva* sp. using existing models by Coffaro and Bocci (1997), Bocci et al. (1997) and Solidoro et al. (1997ab). To deal with the ability of seagrass to survive at low nutrient conditions, we have also included the dynamics of inorganic nitrogen (nitrates and ammonium) in the water column and in the sediments (Chapelle, 1995). A simple phytoplankton model (Plus et al., 2003) has been also incorporated in the main model.

The integrated model is able to simulate successions of dominance states, with different resilience characteristics according with the conceptual scheme. Regime shifts are found when changing nutrient input, temperature and light intensity forcing functions. Finally, a re-interpretation in terms of sensitivity to initial and operating values is discussed for mesocosm experiments.

## 2. METHODS

### 2.1. Model Formulation

The model is based on previous existing and validated models developed for Mediterranean coastal lagoons, i.e. Venice lagoon (Italy) and Etang de Thau (France). This approach was chosen because it would allow the flexibility of analysing different scenarios for these types of ecosystems which are subjected to strong anthropogenic pressures. In addition, previously validated models can offer more robust results than a "*de novo*" approach when there is no experimental data adequate for their validation.

- <u>*Zostera marina* model</u>

The *Zostera marina* sub-model is based on the model described in Bocci *et al*. (1997) and Coffaro and Bocci (1997). State variables in this sub-model are: *Zs* (shoot biomass, gdw m$^{-2}$), *Zr* (rhizome-root biomass, gdw m$^{-2}$) and *Nz* (internal nitrogen quota, mg N gdw$^{-1}$). *Zostera* growth is described as follows:

$$\frac{dZs}{dt} = (growth_z - trans - respiration_s) \cdot Zs \qquad (1)$$

$$\frac{dZr}{dt} = trans \cdot Zs - respiration_r \cdot Zr \qquad (2)$$

$$\frac{dNz}{dt} = uptake_z - growth_z \cdot Nz \qquad (3)$$

The influence of the limiting factors on *Zostera* growth is described with a multiplicative formulation:

$$growth_z = \mu_{max}^z \cdot f_1(I) \cdot f_2(T) \cdot f_3(Nz) \cdot f_4(Zs) \cdot f_5(NO_3^-) \qquad (4)$$

The functional forms as well as the parameters of the model are described in Table 1. The $f_5$ term has been introduced in the Bocci et al. (1997) model to take into account that water-column nitrate enrichment promotes decline of *Zostera marina* independently of algal light attenuation. According to Burkholder et al. (1992) this is probably due to internal imbalances in nutrient supply ratios.

In this formulation, the growth of rhizome depends on the translocation of photosynthetic products from leaves to below-ground parts of the plant. This translocation is proportional to the rate of growth:

$$trans = K_{trans} \cdot growth_z \qquad (5)$$

The parameter $K_{trans}$ was estimated by Olensen and Sand-Jensen (1993) as 25% of the growth, i.e. $K_{trans}=0.25$. Shoot biomass losses are expressed as a function of shoot respiration rate at 20 °C, $SR_{20}$, corrected by the actual temperature:

$$respiration_s = SR_{20} \cdot f_s(T) \qquad (6)$$

where

$$f_s(T) = 0.098 + \exp(-4.690 + 0.2317 \cdot T) \qquad (7)$$

Following a similar approach, rhizome-root biomass loss processes are considered as a function of a respiration coefficient, $RR_{20}$, with a temperature correction:

$$respiration_r = RR_{20} \cdot f_s(T) \qquad (8)$$

Internal nitrogen quota in *Zostera* has been modelled as a function of nitrogen uptake. Shoots can uptake nitrates and ammonium, whereas the rhizome-root can only uptake ammonium.

$$uptake_z = (uptake_s + uptake_r) \cdot f_u(Nz) \qquad (9)$$

$$uptake_s = uptake_s^{NH_4^+} + uptake_s^{NO_3^-} \qquad (10)$$

$$uptake_s^{NH_4^+} = V_{max_s}^{NH_4^+} \frac{[NH_4^+]}{[NH_4^+] + K_s^{NH_4^+}} \qquad (11)$$

$$uptake_s^{NO_3^-} = V_{\max_s}^{NO_3^-} \frac{[NO_3^-]}{[NO_3^-] + K_s^{NO_3^-}} \tag{12}$$

$$uptake_r = V_{\max_r}^{NH_4^+} \frac{[NH_4^+]_s}{[NH_4^+]_s + K_r^{NH_4^+}} \tag{13}$$

$$f_u(Nz) = \frac{Nz_{\max} - Nz}{Nz_{\max} - Nz_{\min}} \tag{14}$$

The values of parameters are summarised in Table 1.

- *Ulva rigida* model

The *Ulva rigida* sub-model is based on the model described in Solidoro *et al*. (1997a,b). State variables in this sub-model are: $U$ (*Ulva* biomass, gdw l$^{-1}$) and $Nu$ (internal nitrogen quota, mg N gdw$^{-1}$). The model can be written as:

$$\frac{dU}{dt} = (growth_u - death_u) \cdot U \tag{15}$$

The influence of the limiting factors on *Ulva* growth was described with a multiplicative formulation:

$$growth_u = \mu_{\max}^u \cdot g_1(I) \cdot g_2(T) \cdot g_3(Nu) \tag{16}$$

The functional forms of the algae model are described in Table 2. As we do not consider oxygen dynamics explicitly, the mortality in this model does not follow Solidoro *et al*. (1997a,b) model. In this case, the mortality term has been expressed as a simple constant and a density dependent function:

$$death_u = k_d^u + k_t^u \cdot \exp\left[-\left(\frac{U - U_{\max}}{U_{width}}\right)^2\right] \tag{17}$$

Like *Zostera marina*, *Ulva* is able to store nitrogen, therefore Solidoro *et al*. (1997a,b) introduced the tissue concentration of this element ($Nu$) as a separated state variable. Its dynamics can be expressed as:

$$\frac{dNu}{dt} = uptake_u - growth_u \cdot Nu \tag{18}$$

The specific uptake rate of nitrogen depends on the chemical form available and on the level $Nu$ of nitrogen tissue concentration. Hence, *uptake* can be written as

$$uptake_u = (uptake_u^{NH_4^+} + uptake_u^{NO_3^-}) \cdot f_u(Nu) \qquad (19)$$

whereas

$$uptake_u^{NH_4^+} = V_{max_u}^{NH_4^+} \frac{[NH_4^+]}{[NH_4^+] + K_u^{NH_4^+}} \qquad (20)$$

$$uptake_u^{NO_3^-} = V_{max_u}^{NO_3^-} \frac{[NO_3^-]}{[NO_3^-] + K_u^{NO_3^-}} \qquad (21)$$

$$f_u(Nu) = \frac{Nu_{max} - Nu}{Nu_{max} - Nu_{min}} \qquad (22)$$

- Phytoplankton model

A simple phytoplankton module was introduced in the model. The module, developed for Etang de Thau (France), has been adapted from Plus et al. (2003). Phytoplankton will compete for nutrients in the water column and will have a shadowing effect - less pronounced than that of *Ulva*- on benthic vegetation.

$$\frac{dP}{dt} = (growth_P - death_P) \cdot P \qquad (23)$$

The influence of the limiting factors on phytoplanton growth was described with a multiplicative formulation (Plus et al., 2003):

$$growth_P = \mu_{max}^P \cdot h_1(I) \cdot h_2(T) \cdot h_3(N) \qquad (24)$$

whereas mortality is also described as a function of temperature:

$$death_P = m_0 \cdot e^{\varepsilon \cdot T} \qquad (25)$$

As we do not explicitly consider zooplankton grazing explicitly, the mortality function in this model has been changed accordingly. Phytoplankton nutrient uptake can be expressed as a function of the nutrient limitation expression and phytoplankton biomass as in Plus et al., (2003). The functional forms of the phytoplankton growth model as well as the main parameters are described in Table 3.

- Dissolved Inorganic Nitrogen (DIN) model

To model the competition between *Zostera* and *Ulva* it is necessary to include nutrient consumption. The nutrients included are nitrogen in the oxidised and reduced forms. Furthermore, in shallow water bodies, sediments play a fundamental role in the nutrient dynamics and in this case *Zostera* is able to uptake ammonium from sediments (Coffaro and Bocci, 1997). For these reasons, the dynamics of DIN in sediments has been introduced as well. The model, adapted from Chapelle (1995), can be written as:

*a/ Dissolved Inorganic Nitrogen (DIN) in the water column*

$$\frac{d[NO_3^-]}{dt} = Nitrif_w - uptake^{NO_3^-} + \frac{Q_{NO_3^-}^{diffusion}}{watervol} + Q_{NO_3^-}^{input} - Q_{NO_3^-}^{output} \quad (26)$$

$$\frac{d[NH_4^+]}{dt} = -Nitrif_W - uptake^{NH_4^+} + \frac{Q_{NH_4^+}^{diffusion}}{watervol} + Q_{NH_4^+}^{input} - Q_{NH_4^+}^{output} \quad (27)$$

Nitrification rates in the water column are functions of water temperature and oxygen concentration, and they can be expressed as:

$$Nnitrif_W = k_{nit} \cdot \exp(k_t \cdot T) \cdot [NH_4^+] \quad (28)$$

$[NO_3^-]$ uptake (mmol N·m$^{-3}$·h$^{-1}$) can be divided into *Ulva*, Zostera and phytoplankton uptake:

$$uptake^{NO_3^-} = \alpha_1 \cdot \left(uptake_u^{NO_3^-} + uptake_z^{NO_3^-}\right) + uptake_p^{NO_3^-} \quad (29)$$

whereas $\alpha_1$ is a conversion factor to pass from mg N to mmol N.

$[NH_4^+]$ uptake (mmol N·m$^{-3}$·h$^{-1}$) can also be divided into *Ulva,* Zostera and phytoplankton uptake:

$$uptake^{NH_4^+} = \alpha_1 \cdot \left(uptake_u^{NH_4^+} + uptake_z^{NH_4^+}\right) + uptake_p^{NH_4^+} \quad (30)$$

At the interface between the water column and the interstitial water, diffusion is responsible for $[NO_3^-]$ and $[NH_4^+]$ fluxes (mmol N·m$^{-3}$·h$^{-1}$). These fluxes can be represented as:

$$Q_{NO_3^-}^{diffusion} = D_{NO_3^-} \frac{A \cdot \beta}{z_s}([NO_3^-]_s - [NO_3^-]) \quad (31)$$

$$Q_{NH_4^+}^{diffusion} = D_{NH_4^+} \frac{A \cdot \beta}{z_s}([NH_4^+]_s - [NH_4^+]) \quad (32)$$

whereas $D_X$ are the sediment diffusion coefficients (m$^2$ h$^{-1}$), $A$ is the exchange area (1 m$^2$), $z_s$ is the distance between the centres of the water and sediment layers, and $\beta$ is the sediment layer porosity (Chapelle, 1995).

This DIN submodel behaves as a CSTR (Continuous Stirred Tank Reactor), where the forcing is given by the fluxes of nutrients. For example, for nitrate it can be written:

$$Q_{NO_3^-}^{input} = \frac{F \cdot [NO_3^-]_{initial}}{V} \tag{33}$$

$$Q_{NO_3^-}^{output} = \frac{F \cdot [NO_3^-]}{V} \tag{34}$$

where $F$ refers to the water flow (m³/h), $V$ to the total volume (1 m³) and $[NO_3^-]_{initial}$ is the initial concentration of nitrate that enters into the system.

*b/ DIN in the sediments*

$$\frac{d[NO_3^-]_s}{dt} = Nitrif_s - Ndenit - \frac{Q_{NO_3^-}^{diffusion}}{interstvol} \tag{35}$$

$$\frac{d[NH_4^+]_s}{dt} = (1 - \alpha_{denit})Ndenit - Nitrif_s - uptake_r - \frac{Q_{NH_4^+}^{diffusion}}{interstvol} \tag{36}$$

Nitrification in the sediments, $Nitrif_s$, can be described as a first order process in ammonium concentration at the sediment:

$$Nnitrif_s = k_{nit} \cdot f_1(T) \cdot f_2(O) \cdot [NH_4^+]_s \tag{37}$$

whereas nitrate reduction can be expressed as a first order process in nitrate concentration at the sediment:

$$Ndenit = k_{denit} \cdot f_1(T) \cdot f_3(O) \cdot [NO_3^-]_s \tag{38}$$

Nitrogen mineralization has not been taken into account in this model. Oxygen concentration is considered constant. The values of parameters taken form Chapelle (1995) are summarised in Table 4.

**2.2. Forcing functions and parameters**

In all the runs, the model has been forced by imposing temperature and solar radiation sinusoidal forcing, which have the following form:

$$T = A_T \cdot \sin\left(2 \cdot \pi\left(\frac{t - 114.74}{365}\right)\right) + T_m \tag{39}$$

$$I = A_I \cdot \left[\sin\left(2\cdot\pi\left(\frac{t-90}{365}\right)\right)+1\right] + I_m \qquad (40)$$

Parameters, amplitude and mean value, were adjusted using meteorological data from several Mediterranean stations, but, in any case, their influence is going to be analyzed.

Nutrient inputs and flows have been maintained constant during each simulation run. This is not typical under natural conditions where nutrient loadings delivered to coastal systems undergo seasonal variations due to rainfall regimes. In addition in the Mediterranean climate region, nutrient loadings to coastal marine systems can attain short term peaks following heavy rainfall events (Plus et al., 2006). Furthermore, oxygen concentrations were set constant at 8.0 g m$^{-3}$ during all simulations, whilst in transitional water ecosystems they undergo daily and seasonal variations from supersaturation to anoxia (Viaroli and Christian, 2003).

**2.3. Assessment of mesocosm data**

Mechanistic experiments dealing with phanerogams-macroalgae-phytoplankton competition were carried out using mesocosms under controlled conditions (Taylor et al., 1999; Nixon et al., 2001). In this work we have re-assessed the mesocosms experiments reported by Taylor et al. (1999). In these experiments (five settings, each with two replicates: Control, C, Low, L, Medium, M, High, H and Very High, VH) enrichment with ammonium and phosphate at several levels was performed and results monitored from April to September.

The following assumptions were made to simulate these experiments:

- Light intensity was assumed to be not a limiting factor for any of the three taxa.

- Temperature was simulated using a sinusoidal function as in Eq. (39) with $T_m$=10.5 and $A_T$=10.1 °C, respectively.

- Dissolved inorganic phosphorous (DIP) was not considered.

- Dissolved inorganic nitrogen (DIN) was equally partitioned between nitrates and ammonium in the background concentration.

- Initial conditions of *Zostera* above-ground biomass were taken constant at 50 gdw m$^{-2}$ whereas the influence of initial conditions of *Ulva* and phytoplankton was assessed.

- A constant flow, $F=5.3 \cdot 10^{-3}$ m$^3$ h$^{-1}$ was assumed during all the experiment, as well as a constant concentration input of nitrate, $[NO_3^-]_{input}$ =2.4 mmol m$^{-3}$ and ammonium, $[NH_4^+]_{input}$ = 2.4, 20.6, 38.7, 75.1, 148.0 mmol m$^{-3}$, for the different experimental conditions (C, L, M, H and VH).

## 3. RESULTS

### 3.1. Competition between *Zostera* and *Ulva*

The first set of simulations was run excluding phytoplankton, to compare with field observations for which no phytoplankton data was mentioned. At low DIN input concentrations (5 mmol m$^{-3}$) *Zostera* survives and *Ulva* disappears (Fig. 1). In addition, due to the relatively high flow of DIN into the system, nitrogen is not completely depleted and the dynamics in the water column and in the sediments are tightly coupled. However, there is a certain transient period of few years before the limit cycle is reached, during which both vegetation types coexist.

The contrary effect, i.e. dominance by *Ulva*, may be observed at high input DIN concentrations (50 mmol m$^{-3}$, see Fig. 2). Keeping high nutrient loads, *Zostera* will disappear after a few years while *Ulva* will tend to prevail.

In order to analyze the effects of DIN inputs in the *Zostera-Ulva* competition model, we have run the model for a set of flow conditions with the same forcing. Figures 3 and 4 present the results in terms of average biomass over the year. *Zostera* dominates the regions with low DIN concentrations whereas the opposite applies to *Ulva*. In addition, due to the fact that in the *Zostera* model the rhizome-root is assumed to only uptake ammonium (Bocci et al, 1997), there is an asymmetry concerning the effects ammonium and nitrate in the figures. Since we consider explicitly the DIN dynamics, the results represented in Figs. 3 and 4 will change as a function of the flow ($F$, m$^3$·h$^{-1}$), assuming the same initial concentrations of nutrients. Likely, at low flows depletion of DIN may occur in the water column as well as in the sediments during the periods of maximum growth. This affects the dynamics in the system and, consequently, the competition between the two taxa. In order to highlight these differences, we have plotted the results obtained with 0.1 and 0.01 m$^3$·h$^{-1}$ flows.

To verify the sensitivity of the competition in relation to changes in temperature, several simulations were set up, with the same conditions as in Fig.1, but with average temperature increased from 0.2 to 2 °C. Results obtained for a temperature increase of 1 ºC after the fifth year are presented in Figure 5. In this case, the outcome is the opposite as in Fig. 1 with *Ulva* dominating the competition. Dynamics and timing of the regime shift are not a simple function of the temperature increase, as shifts have been observed in all the temperature ranges studied depending on the initial and forcing conditions.

The model was also run changing mean temperatures ($T_m$) and annual temperatures ($A_T$) ranges, Eq. (39). Results are presented in Fig. 6, showing that an increase of both parameters tends to favor *Ulva* growth, even in environments with low nutrient concentrations.

Finally the results of the model were analyzed as a function of the incident light. A series of simulation were run by modifying the average light intensity ($I_m$) and its annual range ($A_I$), see Eq. (40). From the results presented in Fig.7 it can be inferred that *Zostera* is adapted to narrower light ranges while *Ulva* seems able to cope with high variable light regimes (Dahlgreen and Kautsky, 2004). However, there is a certain realm of lighting conditions within which *Zostera* dominates even at high DIN concentrations. All simulated results showed that the system was in a transient and the final limit cycle was reached after a few years.

**3.2. The influence of phytoplankton on competition between *Zostera* and *Ulva***

The partitioning of primary production among the different taxa was analyzed under different set of conditions as a function of DIN inputs with high (F=0.1 m$^3$ h$^{-1}$) and low flows (F=0.01 m$^3$ h$^{-1}$), see Fig. 8. Overall, phytoplankton was able to compete with *Ulva* for nutrients in the water column, thus favouring *Zostera* due to its lower shadowing effect. At high DIN loadings phytoplankton outcompeted both *Ulva* and *Zostera*, thus becoming the dominant group. This is due to its higher maximum growth rate (0.021 h$^{-1}$) compared to *Ulva* (0.017 h$^{-1}$) and *Zostera* (0.0025 h$^{-1}$) when no nutrient, temperature or light limitation exists.

**3.3. Assessment of mesocosm experiments**

The model has been used to simulate the mesocosm experiments which tested competition between *Zostera marina*, *Ulva lactuca* and phytoplankton under several nutrient enrichment conditions (Taylor et al., 1999; Nixon et al., 2001). The authors concluded that no significant effect of loading could be detected for *Zostera marina*, epiphytic material, drift macroalgae or for all plant components combined. This contradictory result could be due to several reasons; therefore in this work we have tried to assess two: sensitivity to initial conditions and transient behaviour.

Concerning the sensitivity to initial conditions, results obtained from two identical runs of the mesocosm experiments, but with different initial biomass of *Ulva* and phytoplankton are reported in Figures 9 and 10, as an example. In the first case, *Zostera* biomasses increased steadily with nutrient enrichment, from C to M and decreased from M and VH. In parallel, *Ulva* and phytoplankton increased with nutrient enrichment from C to VH (Fig. 9). However, in the second case (Fig. 10), even though *Ulva* and phytoplankton behaved in a similar way but with delayed dynamics and with different values, *Zostera* showed a different behaviour with higher biomasses at higher concentrations.

## 4. DISCUSSION

The simulated results from the competition between *Zostera* and *Ulva* are in agreement with field observations. For example, DIN concentrations around 5 mmol m$^{-3}$ are typical from Etang de Thau (France), which is covered by *Zostera* meadows; whereas high DIN concentrations ~ 50 mmol m$^{-3}$ have occurred during some years in Sacca di Goro. (Italy), which is dominated by *Ulva*. Similar observations have also been reported by Nelson et al. (2003) with *Ulva* starting to appear at DIN concentrations higher than 18.0 mmol m$^{-3}$. However, the regime shift would be more abrupt since, with such high *Ulva* biomasses, *Zostera* would disappear not only due to nutrient competition, but also due to alteration of sediment biogeochemistry (Holmer et al., 2003) and anoxic crises triggered off by the biomass decomposition (Zaldívar et al., 2003).

Biomasses and *Ulva-Zostera* competition are more correlated with DIN loads than with mean DIN concentrations, since with high growth rates nutrients become depleted. This is one of the reasons why field observations are difficult to use for defining a regime shift value.

Simulation outcomes evidenced that system responses to DIN loadings are complex depending on multiple parameters. For example, environmental conditions, such as temperature and light intensity, seem to play an important role in controlling the competition between benthic and pelagic species. Therefore, attempts to develop a simple nutrient scale for detecting regime shift in benthic vegetation seems not possible. This is probably one of the reasons why experimental observations and mesocosm data do not provide a clear threshold/range of values for regime shifts.

The results of simulations considering the influence of temperature and light intensity can be informative on climatic conditions and depths at which *Zoostera* is able to grow when competing with *Ulva* by providing plausible values at which regime shifts will occur. The simulations can also help the debate on how changes in incident light's spectrum and intensity would affect the benthic vegetation.

The simulation of mesocosm experiments highlighted that the model of benthic vegetation is very sensitivity to initial (biomass concentrations) and operating (temperature, light intensity, DIN flows) conditions. Such result is in agreement with high variability detected in data, where biomasses differed by a factor of two between the experimental replicates (Taylor et al., 1999). In addition, transient regimes last longer (years) than the duration of the experiments (months); therefore, results could be not adequate to demonstrate the effects of nutrient enrichment on plant competition. Differences could be amplified by manipulations, e.g. when setting mesocosms with sediment transfer and phanerogams transplanting.

## 5. CONCLUSIONS

In this work a competition model has been developed with the aim of analysing the succession of primary producer communities in coastal shallow ecosystems and identifying possible nutrient thresholds which cause shifts between alternative stable states.

The integrated model is able to simulate succession of dominance states, with different resilience characteristics according with the conceptual scheme that sees floating macroalgae as the optimal competitors for light, and submerged phanerogams as most efficient in recovering and storing nutrients from the sediments and from the water column. The shift from phanerogams to macrolgae, and finally to phytoplankton dominated communities, conformed to the general theory of succession in coastal lagoons (Viaroli et al., 2008). Field observations support the view that in nutrient poor ecosystems, rizophytes dominate until they are not limited by light penetration (depth effect) or by turbidity and shading by floating vegetation and phytoplankton (Dahlgreen and Kautsky, 2004). Increasing loading rates support the development of macroalgae, whilst high loaded water masses become dominated by phytoplankton.

Regime shifts are found when changing the input of nutrients, but also, model simulations were sensitive to environmental forcing: temperature and light.

Overall, model runs evidenced a clear tendency towards a shift from seagrass to macroalgae under increasing temperatures. However, it is expected that the occurrence and severity of the shifts will be site specific depending on local conditions and past history. These results point out that one of the possible outcomes of an average air temperature increase will be the increase in macroalgae and decrease in benthic vegetation. However, the results of the analysis of a competition model between two species are not sufficient to sustain this point.

The model shows a high sensitivity to initial conditions as well as to forcing parameters, but this effect is also observed in mesocom experiments (Taylor et al., 1999). Furthermore, model simulations show that, when initial conditions do not correspond to steady state conditions, seagrasses communities require time periods to attain steady state, which usually are longer than the duration of the mesocosm experiments. In addition, ecosystems, as other non-linear dynamical systems, are sensitive to initial conditions and even a small difference may drive the system to a completely different position in state space after a certain time (Pahl-Wostl, 1995). Our results suggest that this is probably the main reason behind the high variability found by Taylor et al. (1999) in their experiments, which did not allow finding a clear correlation between nutrient increase and regime shifts.

In its present form, the model does not take into consideration several important aspects such as hydrodynamics, buffering capacity (De Wit et al., 2001; Viaroli et al., 2008), salinity, organic nutrient, oxygen, zooplankton and bacteria as well as interactions between *Ulva* and aquaculture activities. In order to develop a more realistic assessment of regime shifts in terms of range of concentrations and temperature, we plan to consider a real case study in which the studied taxa coexist. Future efforts will aim to implement the competition model using a 3D hydrodynamic approach such as COHERENS (Luyten et al., 1999) for Thau lagoon (France).

**Acknowledgements.** This research has been partially supported by the THRESHOLDS (FP6 Integrated Project Thresholds of Environmental Sustainability, contract nº 003933) and the DITTY (FP5 Energy, Environment and Sustainable Development Programme EVK3-CT-2002-00084) projects. F.S.B. and E.H.-G. acknowledge also support from Spanish MEC and FEDER through the project FISICOS (FIS2007-60327).

# TABLES AND FIGURES

Table 1. Parameters and computed quantities used in the *Zostera marina* model from Bocci *et al.* (1997) and Coffaro and Bocci (1997).

| Parameters, computed quantities | Description | Value |
|---|---|---|
| $\mu_{max}^z$ | Maximum specific growth, | 0.0025 h$^{-1}$ |
| $f_1(I)$ | $f_1(I) = \dfrac{I}{I + K_I^z}$ | |
| $K_I^z$ | Semisaturation constant for light | 500 Kcal m$^{-2}$ d$^{-1}$ |
| | $I = I_0 \exp[-(\varepsilon_w + \varepsilon_u \cdot U)z]$ | |
| $\varepsilon_w$ | Water extinction coefficient | 0.4 m$^{-1}$ |
| $\varepsilon_u$ | Ulva shading coefficient | 40 l gdw$^{-1}$ m$^{-1}$ |
| $f_2(T)$ | $f_2(T) = \exp\left[-\left(\dfrac{T - T_{opt}^z}{T_{width}^z}\right)^2\right]$ | |
| $T_{opt}^z$ | Optimal temperature | 20 °C |
| $T_{width}^z$ | Temperature range, sigmoid width | 3.6 °C |
| $f_3(Nz)$ | $f_3(Nz) = \dfrac{Nz - Nz_{min}}{Nz_{crit} - Nz_{min}}$ | |
| $Nz_{min}$ | Minimum internal nitrogen quota | 5.0 mg N gdw$^{-1}$ |
| $Nz_{max}$ | Maximum internal nitrogen quota | 30.0 mg N gdw$^{-1}$ |
| $Nz_{crit}$ | Critical internal nitrogen quota | 15.0 mg N gdw$^{-1}$ |
| $f_4(Zs)$ | $f_4(Zs) = 1 - \exp\left[-\left(\dfrac{Zs - Zs_{max}}{Zs_{width}}\right)^2\right]$ | |
| $Zs_{max}$ | Maximum shoot biomass | 500 gdw m$^{-2}$ |
| $Zs_{width}$ | Growth dependence on space availability | 5 gdw m$^{-2}$ |
| | $f_5(NO_3^-) = \exp\left[-\left(\dfrac{NO_3^- - NO_{3_{opt}}^-}{NO_{3_{width}}^-}\right)^2\right]$ | |
| $NO_{3_{opt}}^-$ | Optimal nitrate concentration | 5.0 mmol m$^{-3}$ |
| $NO_{3_{width}}^-$ | Nitrate concentration range | 80.0 mmol m$^{-3}$ |
| $SR_{20}$ | Shoot respiration rate at 20 °C | 1.0042·10$^{-3}$ h$^{-1}$ |
| $RR_{20}$ | Rhizome-root respiration rate at 20 °C | 6.25·10$^{-4}$ h$^{-1}$ |
| $V_{max_s}^{NH_4^+}$ | Shoot maximum uptake for $NH_4^+$ | 0.3 mgN gdw$^{-1}$ h$^{-1}$ |
| $K_s^{NH_4^+}$ | Shoot half saturation constant for $NH_4^+$ | 9.29 mmol N m$^{-3}$ |
| $V_{max_s}^{NO_3^-}$ | Shoot maximum uptake for $NO_3^-$ | 0.06 mgN gdw$^{-1}$ h$^{-1}$ |
| $K_s^{NO_3^-}$ | Shoot half saturation constant for $NO_3^-$ | 16.43 mmol N m$^{-3}$ |
| $V_{max_r}^{NH_4^+}$ | Rhizome-root maximum uptake for $NH_4^+$ | 0.02 mgN gdw$^{-1}$ h$^{-1}$ |
| $K_r^{NH_4^+}$ | Rhizome-root half saturation constant for $NH_4^+$ | 5.0 mmol N m$^{-3}$ |

Table 2. Parameters and computed quantities used in the *Ulva* model from Solidoro *et al.* (1997a) and Solidoro *et al.* (1997b).

| Parameters, computed quantities | Description | Value |
|---|---|---|
| $\mu_{max}^u$ | Maximum specific growth, | 0.0167 h$^{-1}$ |
| $g_1(I)$ | $g_1(I) = \dfrac{I}{I + K_I^u}$ | |
| $K_I^u$ | Semisaturation constant for light | 239 Kcal m$^{-2}$ d$^{-1}$ |
| $g_2(T)$ | $g_2(T) = \dfrac{1}{1 + \exp(-\varsigma(T - T_U))}$ | |
| $\varsigma$ | Temperature Coefficient | 0.2 °C$^{-1}$ |
| $T_U$ | Temperature reference | 12.5 °C |
| $g_3(Nu)$ | $g_3(Nu) = \dfrac{Nu - Nu_{min}}{Nu - Nu_{crit}}$ | |
| $Nu_{min}$ | Min. value for N quota | 10.0 mg N/gdw |
| $Nu_{crit}$ | Critical N quota level | 7.0 mg N/gdw |
| $Nu_{max}$ | Max. value for N quota, uptake limitation | 42.0 mg N/gdw |
| $k_d^u$ | Mortality rate | 6.2·10$^{-3}$ h$^{-1}$ |
| $k_t^u$ | Mortality rate due to biomass | 1.0 h$^{-1}$ |
| $U_{max}$ | Maximum *Ulva* biomass | 0.6 gdw l$^{-1}$ |
| $U_{width}$ | Growth dependence on space availability | 0.01 gdw l$^{-1}$ |
| $V_{max_u}^{NH_4^+}$ | Max. specific uptake rate for ammonium | 8.5 mg N gdw$^{-1}$ h$^{-1}$ |
| $V_{max_u}^{NO_3^-}$ | Max. specific uptake rate for nitrate | 0.45 mg N gdw$^{-1}$ h$^{-1}$ |
| $K_u^{NH_4^+}$ | Half-saturation for ammonium | 7.14 mmol/m$^3$ |
| $K_u^{NO_3^-}$ | Half-saturation for nitrate | 3.57 mmol/m$^3$ |

Table 3. Parameters and computed quantities used in the phytoplankton model from Plus *et al.* (2003).

| Parameters, computed quantities | Description | Value |
|---|---|---|
| $\mu_{max}^P$ | Maximum specific growth, | 0.021 h$^{-1}$ |
| $h_1(I)$ | $h_1(I) = \left(1 - e^{-I/I_k}\right)$ | |
| $I_k$ | Saturation constant for light | 620.1 Kcal m$^{-2}$ d$^{-1}$ |
| $h_2(T)$ | $h_2(T) = e^{\varepsilon \cdot T}$ | |
| $\varepsilon$ | Temperature Coefficient | 0.07 °C$^{-1}$ |
| $h_3(N)$ | $\dfrac{[NH_4^+]}{[NH_4^+] + K_N} + \dfrac{[NO_3^-]}{[NO_3^-] + K_N} e^{-\psi \cdot [NH_4^+]}$ | |
| $K_N$ | Half saturation constant for N limitation | 2.0 mmol m$^{-3}$ |
| $\psi$ | Wroblewski inhibition factor | 1.5 m$^3$ mmol$^{-1}$ |
| $m_0$ | Mortality rate at 0 C | 1.15·10$^{-2}$ h$^{-1}$ |

Table 4. Nutrient and sediment parameters, from Chapelle (1995).

| Parameters, computed quantities | Description | Value |
|---|---|---|
| $k_{nit}$ | Nitrification rate at 0° C | 0.0083 h$^{-1}$ |
| $f_1(T)$ | $f_1(T) = \exp[k_T \cdot T]$ | |
| $k_T$ | Temperature increasing rate | 0.07 °C$^{-1}$ |
| $f_2(O)$ | $f_2(O) = \dfrac{[O_2]_s}{K_{NitO} + [O_2]_s}$ | |
| $k_{Nit}O$ | Half-saturation coefficient for O$_2$ limitation of nitrification | 4.0 g/m$^3$ |
| $RUPC$ | *Ulva* stoichiometric ratio | 2.5 mg P/gdw |
| $QPS$ | Photosynthetic ratio | 1.5 |
| $RPHY$ | Phytoplankton respiration rate at 0 C | 2.083.10$^{-3}$ h$^{-1}$ |
| $\psi$ | Stoichiometric ratio | 1450 g O$_2$/gdw |
| $RPS$ | O2 produced/N | 0.212 g O$_2$/mmol |
| $D_{NO}$ | Diffusion coefficient for nitrate in the sediment | 0.00072 m$^2$/h |
| $D_{NH}$ | Diffusion coefficient for ammonium in the sediment | 0.00072 m$^2$/h |
| $A$ | Surface of computationa cell | 1 m$^2$ |
| $z_S$ | Distance between the centre of water cell and sediment layer | 0.51 m |
| *watervol* | Volume of the water cell | 1 m$^3$ |
| *interstvol* | Interstitial water volume for a cell | 0.008 m$^3$ |
| $k_{denit}$ | Denitrification rate at 0° C | 0.0125 h$^{-1}$ |
| $f_3(O)$ | $f_3(O) = 1 - \dfrac{[O_2]_s}{K_{denitO} + [O_2]_s}$ | |
| $[O_2]$ | Oxygen concentrarion | 8 g/m$^3$ |
| $K_{denitO}$ | Half-saturation coefficient for O$_2$ limitation of denitrification | 2.0 g/m$^3$ |
| $\alpha_{denit}$ | percentage of N denitrified into N$_2$ | 0.6 |
| $f_4(O)$ | $f_4(O) = \dfrac{[O_2]_s}{K_{minO} + [O_2]_s}$ | |
| $K_{minO}$ | Half-saturation coefficient for O$_2$ limitation of mineralization | 0.5 g/m$^3$ |

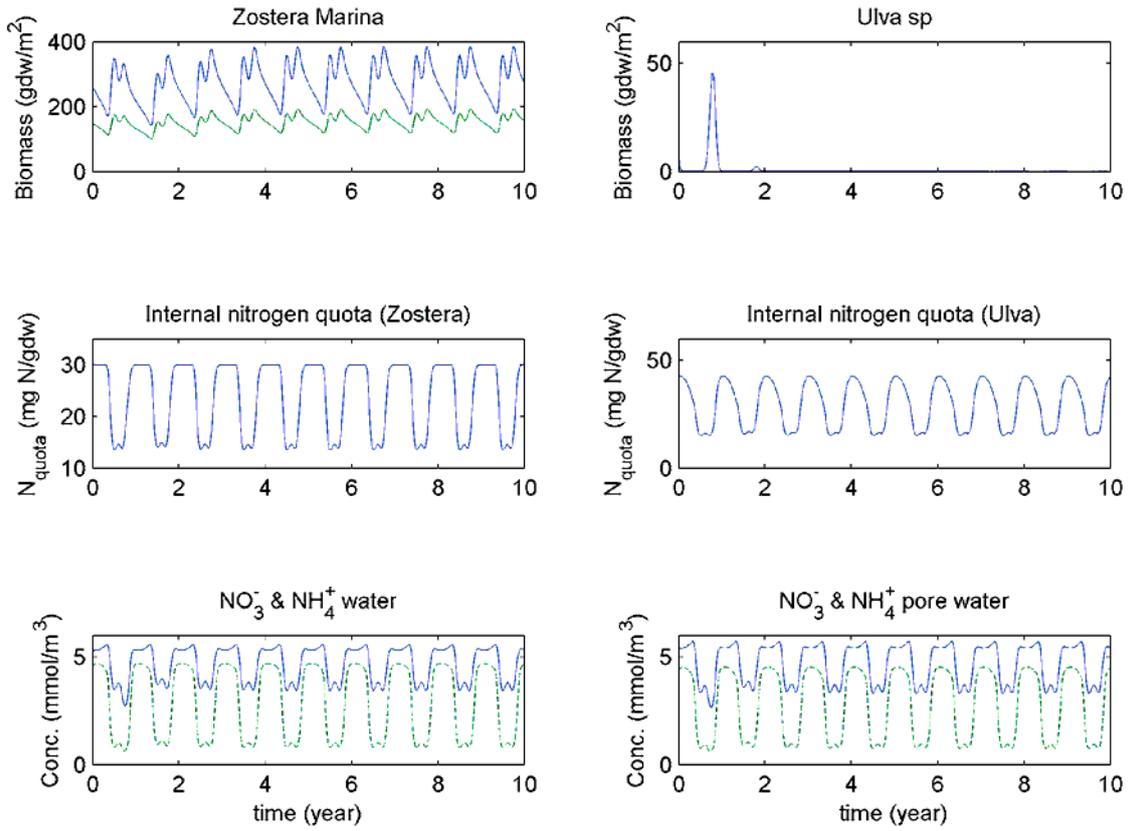

Figure 1. Zostera biomasses $Zs$ (shoot biomass, gdw m$^{-2}$) and $Zr$ (rhizome-root biomass, gdw m$^{-2}$-in green-); Ulva biomasses (gdw m$^{-2}$); Internal nitrogen quotas; DIN concentrations ($NO_3^-$: blue, $NH_4^+$: green) in the water column and in the sediments (pore water). $F$ =0.1 m$^3 \cdot$h$^{-1}$, and $[NO_3^-]_{input} = [NH_4^+]_{input}$ =5 mmol m$^{-3}$ (low nutrient situation).

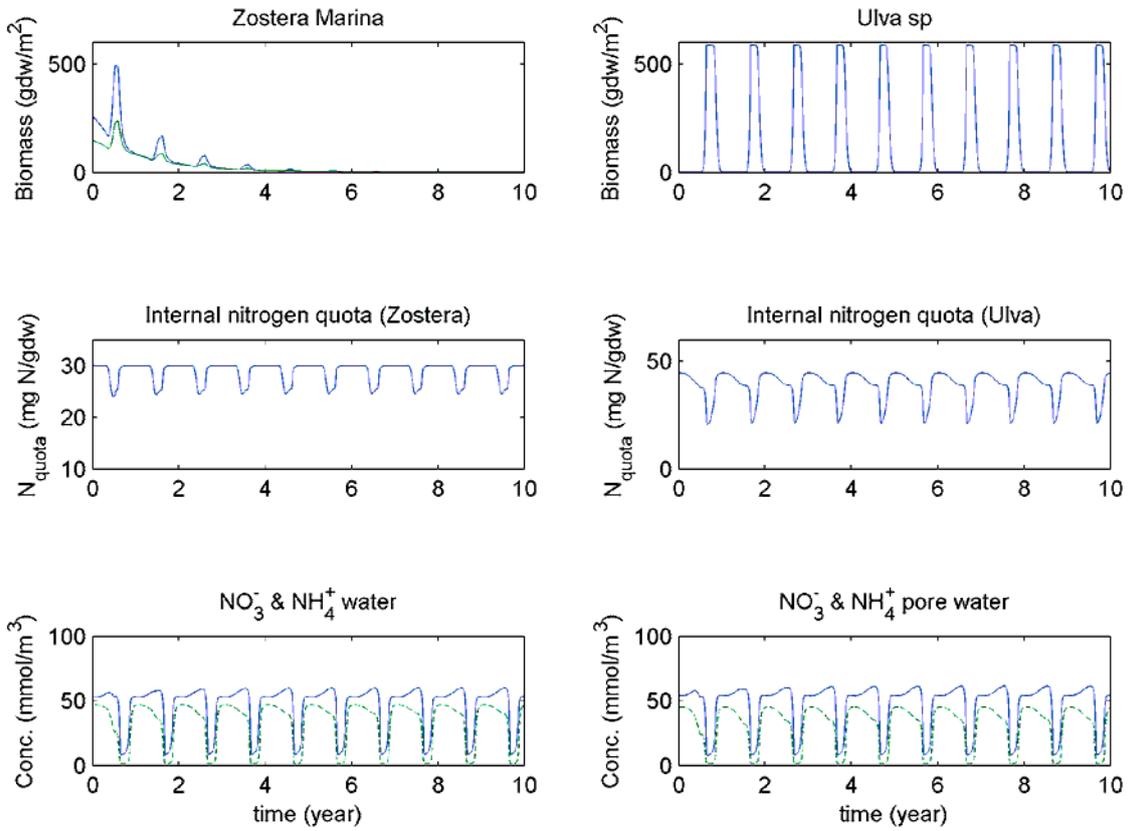

Figure 2. *Zostera* biomasses: *Zs* (shoot biomass, gdw m$^{-2}$) and *Zr* (rhizome-root biomass, gdw m$^{-2}$ -in green-); Ulva biomasses (gdw m$^{-2}$); Internal nitrogen quotas; DIN concentrations ($NO_3^-$: blue, $NH_4^+$: green) in the water column and in the sediments (pore water). $F$ =0.1 m$^3 \cdot$h$^{-1}$, and $[NO_3^-]_{input} = [NH_4^+]_{input}$ =50 mmol m$^{-3}$ (High nutrient situation).

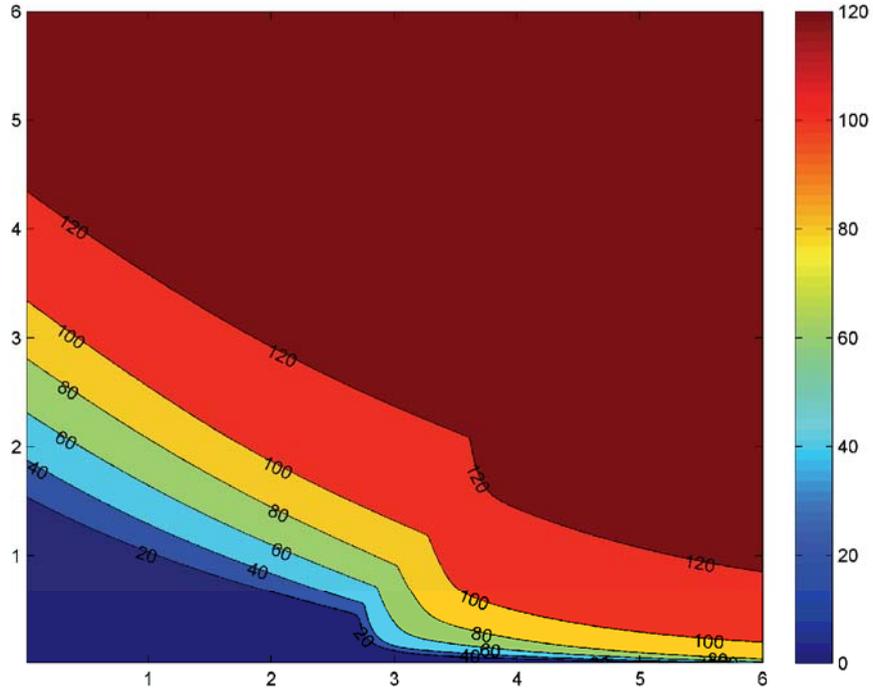

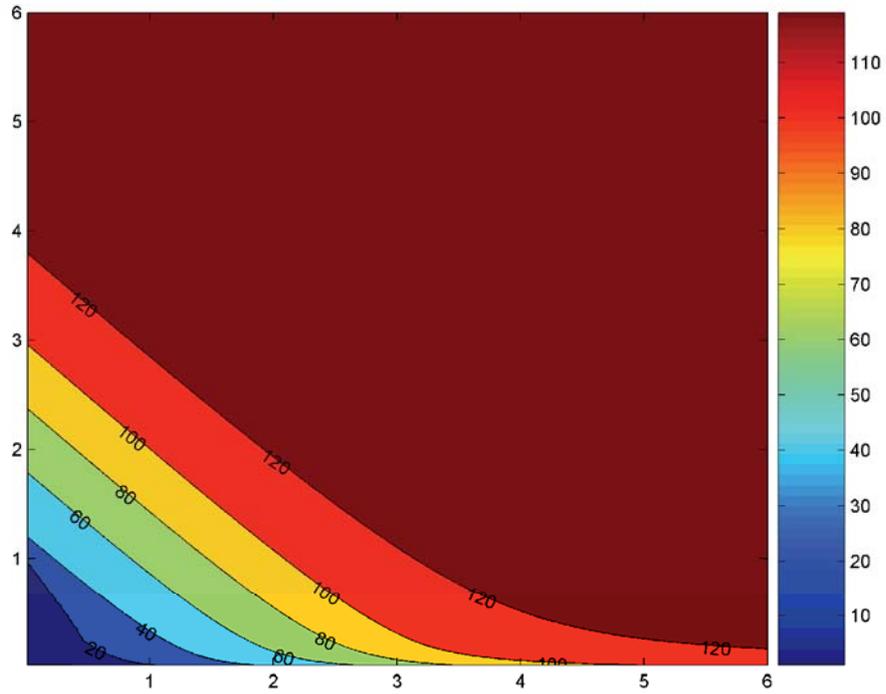

Figure 3. *Ulva* annual mean biomass (gdw m$^{-2}$) as a function of nitrate (x-axes) and ammonium loads (y-axes) in mmol h$^{-1}$. Top: $F$=0.1 m$^3 \cdot$h$^{-1}$; bottom: $F$=0.01 m$^3 \cdot$h$^{-1}$.

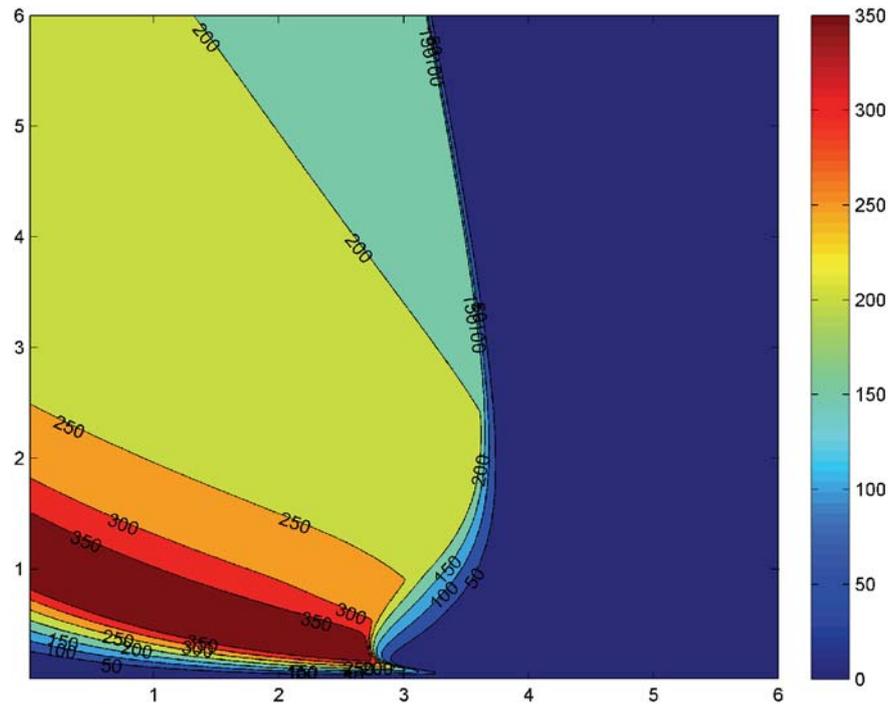

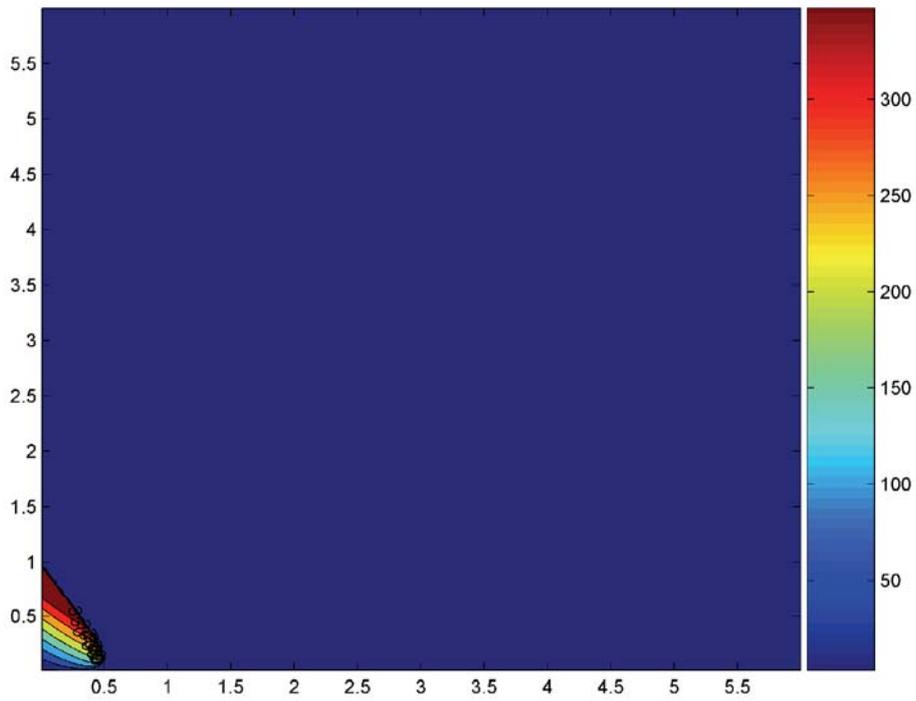

Figure 4. *Zostera* annual mean biomass (gdw m$^{-2}$) as a function of nitrate (x-axes) and ammonium (y-axes) loads in mmol h$^{-1}$. Top: $F$=0.1 m$^3 \cdot$h$^{-1}$; bottom: $F$=0.01 m$^3 \cdot$h$^{-1}$.

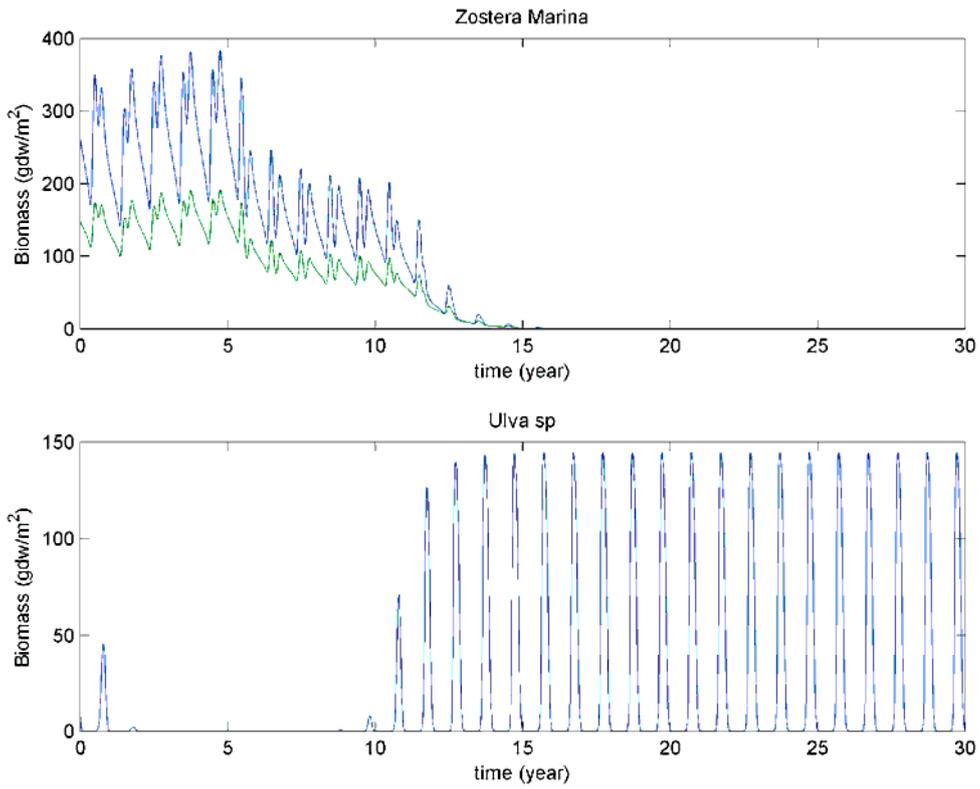

Figure 5. *Zostera* biomasses: *Zs* (shoot biomass, gdw m$^{-2}$, blue) and *Zr* (rhizome-root biomass, gdw m$^{-2}$, green) and *Ulva* biomasses (gdw m$^{-2}$). Same parameters as in Fig. 1, but after the fifth year of simulation the temperature forcing function increases by 1.0° C.

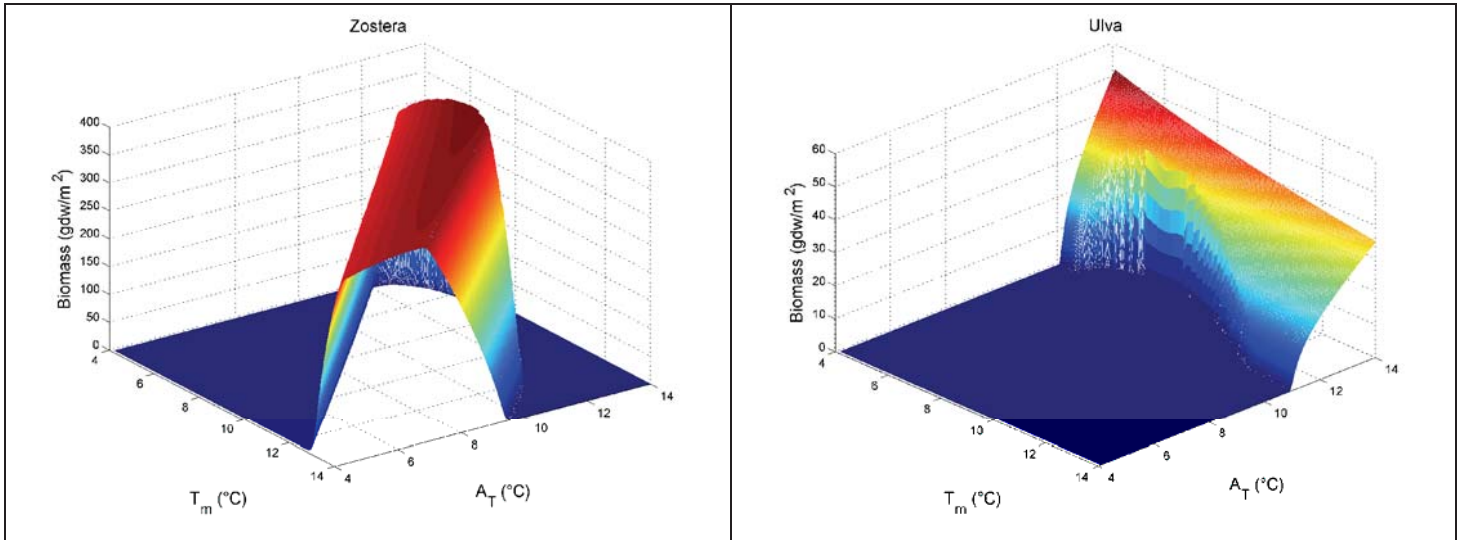

Figure 6. Zostera and Ulva annual mean biomass (gdw m$^{-2}$) as a function of mean temperatures, $T_m$, and its amplitude of annual variation, $A_T$, for the low nutrient regime, $F$=0.1 m$^3 \cdot$h$^{-1}$ and $[NO_3^-]_{input} = [NH_4^+]_{input}$ =5 mmol m$^{-3}$.

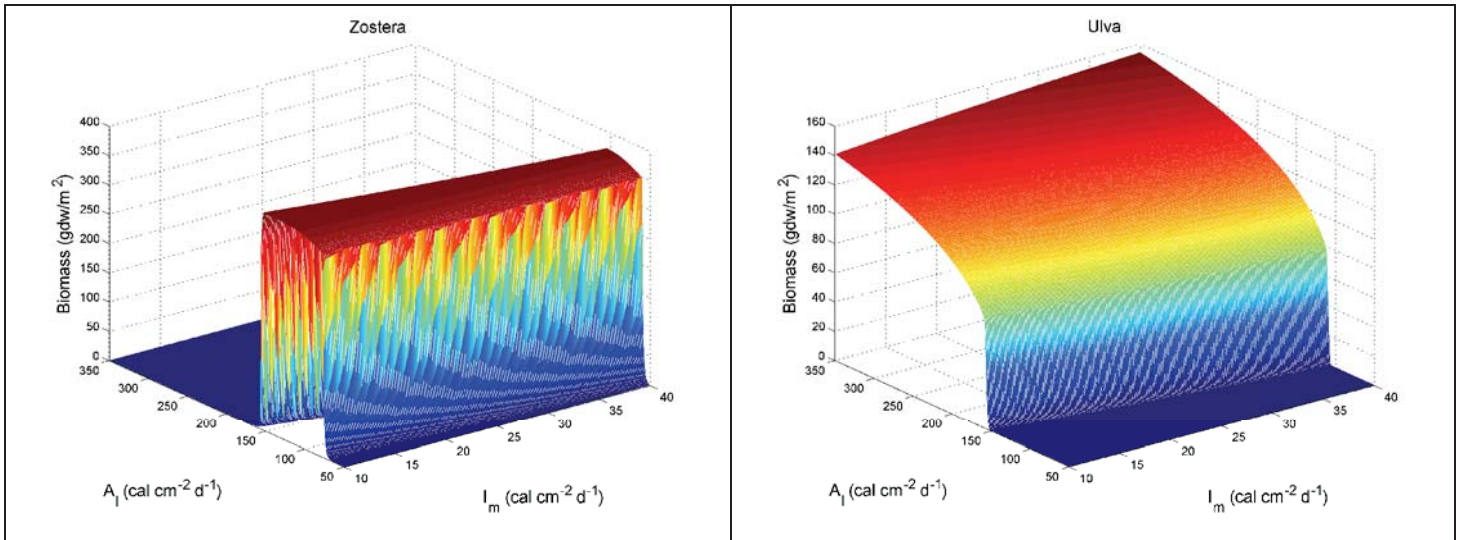

Figure 7. *Zostera* and *Ulva* annual mean biomass (gdw m$^{-2}$) as a function of mean light intensity, $I_m$, and its amplitude of annual variation, $A_I$, for the high nutrient regime, $F$=0.1 m$^3$·h$^{-1}$ and $[NO_3^-]_{input} = [NH_4^+]_{input}$ =50 mmol m$^{-3}$.

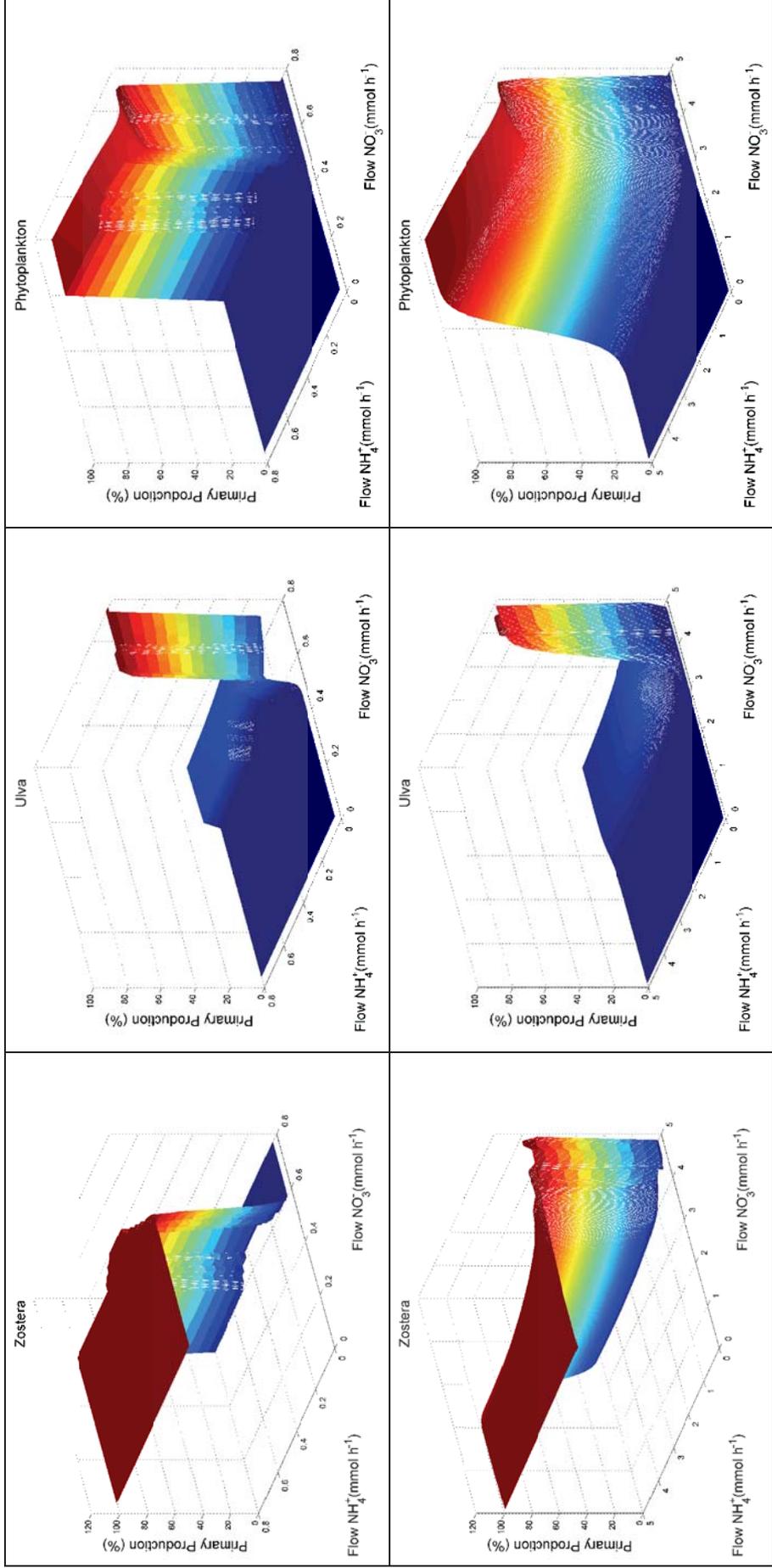

Figure 8. Percentage of primary production for each compartment: *Zostera*, *Ulva* and phytoplankton as a function of nutrient flows. Top: $F$=0.01 m$^3$·h$^{-1}$ ; bottom: $F$=0.1 m$^3$·h$^{-1}$ .

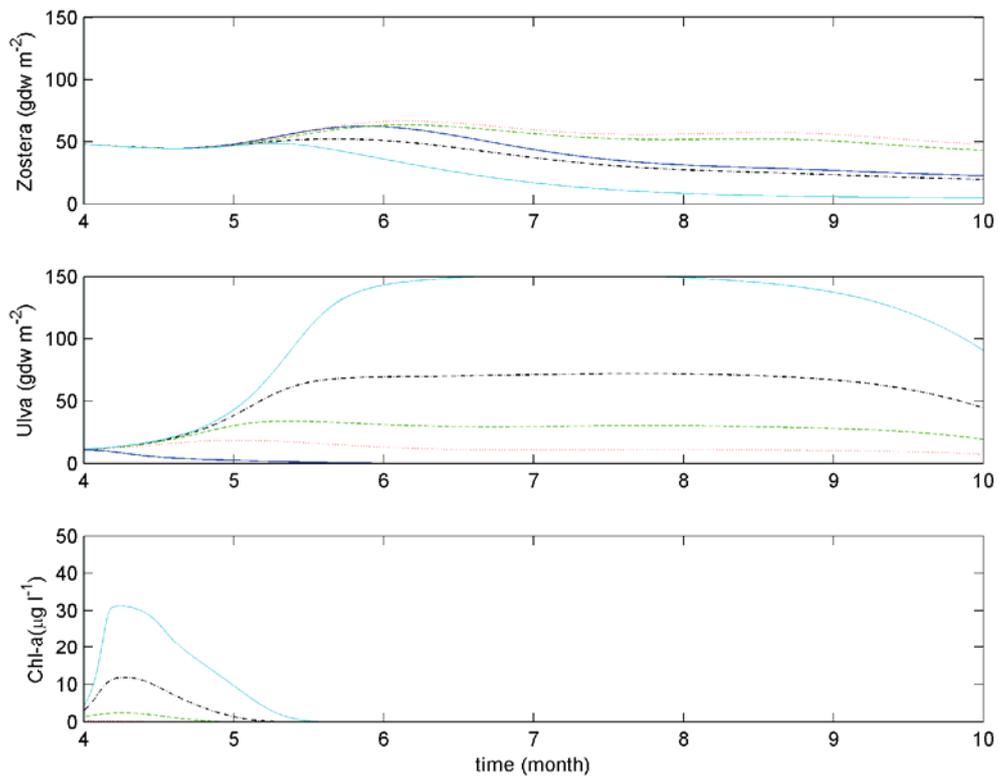

Figure 9. Simulated biomasses of *Zostera*, *Ulva* and phytoplankton during the mesocosm experiments from Taylor et al. (1999). Control (C): continuous blue line; Low (L):dotted red line(L); Medium (M):dashed green line; High (H): dash-dot black line; Very high (VH): continuous cyan line.

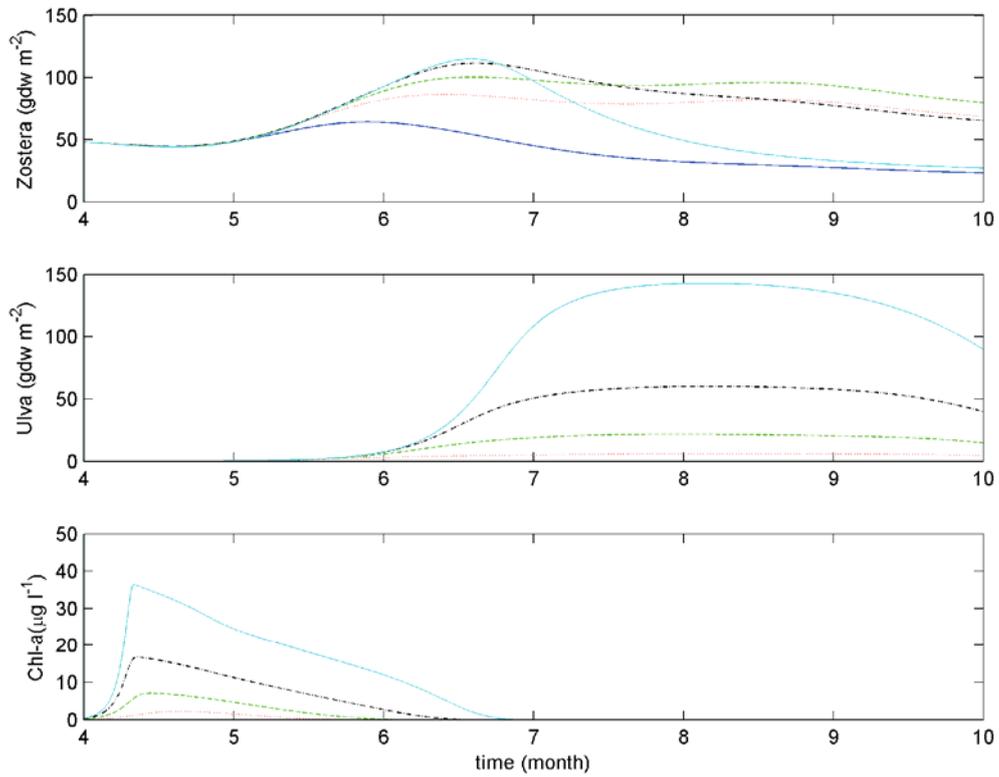

Figure 10. Simulated biomasses of *Zostera*, *Ulva* and phytoplankton during the mesocosm experiments from Taylor et al. (1999). Control (C): continuous blue line; Low (L):dotted red line(L); Medium (M):dashed green line; High (H): dash-dot black line; Very high (VH): continuous cyan line. Conditions as Fig. 9 but the initial conditions of *Ulva* and phytoplankton have been reduced by a factor of ten.